\documentstyle[12pt]{article}

\title{Continuously Measured Systems,\\ 
	Path Integrals and Information\thanks{Published in Intern.  
	J. Theor.  Phys. 37, 273-280 (1998)}} 
\author{Michael B. Mensky\thanks{Email: mensky@sci.lebedev.ru}\\
	P.N.Lebedev Physical Institute\\ Leninsly prosp.43, 117924 
Moscow, Russia}
\date{}

\newcommand{\al}{\alpha}

\newcommand{\be}{\begin{equation}}
\newcommand{\ee}{\end{equation}}
\newcommand{\eq}[1]{(\ref{#1})}
\newcommand{\Eq}{Eq.~\eq}

\newcommand{\ra}{\rangle}
\newcommand{\la}{\langle}

\begin{document}
\maketitle
\begin{abstract}
A continuously measured quantum system may be described by 
restricted path integrals (RPI) or equivalently by non-Hermitian 
Hamiltonians. The measured system is then considered as an open 
system, the influence of the environment being taken into account 
by restricting the path integral or by inclusion of an imaginary 
part in the Hamiltonian. This way of description of measurements 
naturally follows from the Feynman form of quantum mechanics 
without any additional postulates and may be interpreted as an 
information approach to continuous quantum measurements. 
This reveals deep features of quantum physics concerning relations 
between quantum world and its classical appearance. 
\end{abstract}

\section{Introduction}

It is widely believed that quantum mechanics is not closed, and 
that only after adding some form of quantum theory of measurement 
does it become 
a complete and self-sufficient theory. We shall argue that the theory of 
continuous quantum measurements may in fact be considered as a 
natural part of quantum mechanics provided the latter is taken in the 
Feynman path-integral form \cite{Feynman48} including the rules 
for summing up probability amplitudes. The main instruments of the 
resulting theory of continuous measurements are restricted path 
integrals (RPI) and non-Hermitian Hamiltonians \cite{M79,book93}.  
The RPI approach may be regarded as an information approach to 
continuous quantum measurements just as the von Neumann's 
projection postulate presents an information approach to 
instantaneous quantum measurements. 

So-called ``instantaneous" measurements (which are in reality not 
instantaneous, but very short) may be obtained as a limiting case 
of continuous measurements. Therefore, the whole quantum theory 
of measurements may be derived, in the framework of the RPI 
approach, from quantum mechanics in the path-integral form. 
Hence, quantum mechanics may be considered as 
a closed theory. It looks nonclosed only if the 
overidealized concept of an instantaneous measurement is 
considered instead of realistic concept of a continuous 
measurement. 

\section{Continuous quantum measurements and 
restricted path integrals (RPI)} \label{SecRPI}

During recent decades the theory of {continuous quantum 
measurements} has been under thorough investigation 
\cite{M79}-\cite{cont-meas}. 
The interest in this field significantly increased in connection 
with the {quantum Zeno effect} predicted in  \cite{Zeno} and 
experimentally verified in \cite{Itano}. In most cases studying 
continuous quantum measurements was based on particular {models} 
of measuring devices. In contrast, the phenomenological and 
therefore model-independent restricted-path-integrals (RPI) 
approach to continuous measurements has been proposed in 
\cite{M79,Mbk2,book93} (see also \cite{RPIothers}) following the 
idea of R.Feynman \cite{Feynman48}. 

The measured system is considered in the RPI theory of 
measurements as an open system. The back influence of the 
measuring device (environment) onto the measured system is 
taken into account by restricting the path integral. The 
restriction is determined by the information 
which the measurement supplies about the measured system. 
Let us consider the main points of this approach. 

The evolution of a closed quantum system during a time interval 
$T$ is described by the evolution operator $U_T$. 
The matrix element of the operator $U_T$ between the states with 
definite positions is called the propagator and may be expressed 
in the 
form of the Feynman path integral:\footnote{It is convenient for 
our goals to use a phase-space representation of the path 
integral. The variables $q$ and $p$ may be multidimensional.}
\be 
U_T(q'',q')=\la q''|U_T|q'\ra=\int d[p]d[q]\, 
e^{\frac{i}{\hbar}\int_0^T (p\dot q - H(p,q,t))}. 
\label{Feyn}\ee 

If the system with the same dynamical properties (the same 
Hamiltonian) undergoes a continuous (prolonged in time) 
measurement (and therefore is considered as being open,
interacting with a measuring device or environment), its 
evolution may be described \cite{book93} by the set of 
{\em partial evolution operators} $U_T^{\alpha}$ depending 
on the output (readout) $\alpha$ of the 
measurement\footnote{Physically the readout is recorded 
in some way or another in the state of the environment 
(measuring device).} 
$$
|\psi_T^{\alpha}\rangle = U_T^{\alpha} |\psi_0\rangle, \quad
\rho_T^{\alpha} = U_T^{\alpha} \rho_0 \left(
U_T^{\alpha}\right)^{\dagger}.  
$$

The partial propagators are expressed by restricted 
path integrals. 
This means \cite{book93} that the path integral for 
$U_T^{\alpha} $ must be of the form (\ref{Feyn}), but 
restricted according to the information 
given by the measurement readout $\alpha$. The information 
given by $\alpha$ may be described by a weight functional 
$w_\alpha[p,q]$ (positive, with values between 0 and 1) 
so that the partial propagator has to be written as a weighted 
path integral 
\be 
U_T^{\alpha}(q'',q') =
\la q''|U_T^{\alpha}|q'\ra=
\int d[p]d[q]\,w_\alpha[p,q]\, 
e^{\frac{i}{\hbar}\int_0^T (p\dot q - H(p,q,t))}.  
\label{RPIweight}\ee 

The probability density for each $\alpha$ to arise as a 
measurement readout is given \cite{book93} by the trace of the 
density matrix $\rho_T^{\alpha}$, so that the probability for  
$\alpha$ to belong to some set $\cal A$ of readouts is equal to 
\be
{\rm {Prob}}(\alpha\in{\cal A})=
\int_{{\cal A}} d\alpha\, {\rm Tr}\, \rho_T^{\alpha}
\label{prob}\ee
with an appropriate measure $d\alpha$ on the set of readouts. 

All this concerns the situation when the measurement 
readout $\alpha$ is known (selective description of the 
measurement). If the readout is unknown (nonselective 
description), the evolution of the measured system may be 
presented by the density matrix 
\be
\rho_T = \int d\alpha\,  \rho_T^{\alpha}
=\int d\alpha\, U_T^{\alpha} \rho_0 \left( U_T^{\alpha}\right)^{\dagger}.
\label{dens-matr}\ee
The generalized unitarity condition 
$$
\int d\alpha\,  \left( U_T^{\alpha}\right)^{\dagger}\, U_T^{\alpha} =
\bf 1
$$
provides conservation of probabilities. 

In the special case, when monitoring an observable $A=A(p,q,t)$ 
is considered as a continuous measurement,  the measurement 
readout is given by the curve 
$$
[a] = \{a(t)|0\le t \le T\}
$$
characterizing the values of this observable in different time moments. 
If  the square average deflection  
$$
\langle (A-a)^2\rangle_T
= \frac{1}{T}\int_0^T
[ A(t) - a(t) ]^2\,dt 
$$
is taken as a measure of deviation of the observable 
$A(t)=A(p(t),q(t),t)$ from the readout $a(t) $, then the weight 
functional describing the measurement may be 
taken\footnote{The choice of the weight functional determines 
the class of measurements under consideration.} in the 
Gaussian form:
$$
w_{[a]}[p,q] = e^{-\kappa \int_0^T
[ A(t) - a(t) ]^2\,dt }. 
$$
The constant $\kappa$ characterizes the resolution of the 
measurement and may be expressed in terms of the 
``measurement error'' $\Delta a_T$ which is achieved during
 the period $T$ of the measurement: 
$$
\kappa=\frac{1}{T\Delta a_T^2} \mbox{ is constant, hence } 
\Delta a_T^2\sim \frac{1}{T}
$$

The resulting path integral
$$
U_T^{[a]}(q'',q')=\int d[p]d[q]
\exp\left\{ \frac{i}{\hbar} \int_0^T \big(p\dot q
- H\big)dt  - \kappa \int_0^T \big(A(p,q,t)-a(t)\big)^2 dt
\right\}
$$
has the form of a conventional (nonrestricted) Feynman path 
integral (\ref{Feyn}) but with the {\em effective Hamiltonian}
\be
H_{[a]}(p,q,t) = H(p,q,t) - i\kappa\hbar \,\big( A(p,q,t) - a(t)
\big)^2
\label{effectHam}\ee
instead of the original Hamiltonian $H$. 
Therefore, instead of calculating a restricted path integral one 
may solve the Schr\"odinger equation with a non-Hermitian 
effective Hamiltonian: 
\be
\frac{\partial}{\partial t} |\psi_t\rangle
  = \left(-\frac{i}{\hbar} H
  -\kappa \,\Big( A - a(t)\Big) ^2\right)\, |\psi_t\rangle.
\label{eq-gen}\ee

If we solve this equation with the initial wave function $\psi_0$ 
describing the initial state of the measured system, then the 
solution $\psi_T$ in the final time moment represents the state of 
the system after the measurement, under the condition that the 
measurement results in the readout $[a]$. The wave function 
$\psi_T$ obtained in this way has a non-unit norm. If the initial 
wave function is normalized, then the norm of the final wave 
function, according to \Eq{prob}, determines the probability 
density of the measurement output $[a]$: 
\be
P[a]=||\psi_T||^2. 
\label{prob-dens}\ee

We obtain the following scheme of calculation for the {\em 
selective description} of the continuous measurement (when the 
readout is known): 

1.  Choose an arbitrary readout $[a]$ and solve Eq.~(\ref{eq-gen}). 

2. The probability density of $[a]$ is given by Eq.~(\ref{prob-dens}). 

3. The state of the system after the measurement  is $|\psi_T\rangle$. 

\noindent
The {\em nonselective description} of the measurement (if the 
readout is unknown) is given by the 
density matrix $\rho_t$ defined by (\ref{dens-matr})and satisfying  
\cite{master-eq} the equation 
\begin{equation}\label{non-select}
\dot\rho =-\frac{i}{\hbar}[H,\rho]-\frac{\kappa}{2}[A,[A,\rho]]. 
\end{equation}

\section{RPI as an information approach to continuous quantum 
measurements} \label{info}

The description of continuous quantum measurements by restricted 
path integrals (RPI) may be justified in different ways. The most 
direct way \cite{KMNamiot} is an analysis of a composite system 
containing both the measured system and its environment 
(measuring device). Alternatively, one can consider a 
series of instantaneous measurements with the help of 
von Neumann's projection postulate and then go over to 
the continuous measurement as a limit of small time intervals 
between the instantaneous measurements \cite{Mbk2,book93}. 

It is very interesting, however, that in the framework of the 
Feynman version of quantum mechanics the RPI approach needs 
no justification at all. This approach is natural and self-consistent 
in this framework. This is why R.Feynman was able to formulate 
the idea of the RPI approach as a short remark in his paper 
\cite{Feynman48}. Moreover, the RPI approach, naturally 
following from Feynman quantum mechanics, is more general 
than what can be obtained in the limit of a series of instantaneous 
measurements. It describes a wider class of 
continuous and continual (protracted in time and space) 
measurements than those derivable as limits of some 
or other repeated measurements \cite{book93}. 

The reason why the RPI approach follows from the path-integral  
version of quantum mechanics is that the concept of 
{\em probability amplitude} is used in a much more 
comprehensive way in this version. In particular, 
the amplitude 
\be
A[p,q]=e^{\frac{i}{\hbar}\int_0^T (p\dot q - H(p,q,t))}
\label{Apath}\ee
is introduced and physically interpreted as a probability amplitude 
for the system to propagate along a definite path in the phase space. 
If this is accepted, then the usual quantum-mechanical rules for 
amplitudes determine the amplitudes for more complicated events, in 
particular, for propagation of the system between two points of the 
configuration space. If the system is closed and therefore nothing is 
known about the path along which it propagates, all 
amplitudes of the form \eq{Apath} have to be summed up, leading to 
the conventional Feynman integral \eq{Feyn}. If a continuous 
measurement takes place (so that the system is open), one has to 
keep in mind that the measurement supplies some information about 
the evolution of the system. In summing up the amplitudes \Eq{Apath} 
this information must be taken into account. 

If the information given by the measurement can be expressed by a 
weight functional $w_{\al}[p,q]$, then summation of amplitudes 
takes the form of \Eq{RPIweight}. Hence, instead of directly 
postulating partial propagators, we can {\em derive} them from 
the more basic postulates of the path-integral version of 
quantum mechanics. 

This is both interesting and unexpected. It is commonly believed 
that quantum mechanics is not closed, since it does not include any 
theory of measurements. A theory of measurements (for example, 
von Neumann's projection postulate) is customarily appended as 
a necessary counterpart forming, together with quantum mechanics, 
a closed theory. However, this proves to be unnecessary. As 
argued above, the  path-integral formulation of quantum mechanics 
includes also the RPI theory of continuous measurements. A theory 
of instantaneous measurements (including von Neumann's 
postulate) may be then obtained as a limit \cite{Calarco}. 

We see therefore that the seeming necessity to postulate a theory of 
measurement independently of quantum mechanics is only a 
consequence of overidealization. The origin of this necessity is in 
treating a measurement as an instantaneous 
act. An instantaneous measurement appears to be external with  
respect to quantum mechanical (Schr\"odinger) evolution 
and to need special postulates. The 
situation however is radically different if the measurement is 
considered as a temporarilly extended process and quantum mechanics 
is accepted 
in the path-integral form.  Then the measurement may be described in 
a nonseparably integral way with the quantum-mechanical evolution 
\cite{MBohr-sem}. The mathematical apparatus describing both 
counterparts of this unity is given by restricted path 
integrals. 

Restricted path integrals describe the influence of the measuring 
environment on the open (measured) quantum system without an 
explicit model of the environment. Instead, the RPI 
approach needs only very general characteristics of the 
environment. Namely, it is necessary to know what information 
about the evolution of the system is recorded in the state of the 
environment (as a measurement readout). This information 
determines what weight functional has to be used in the 
path integral. Having a restricted path integral, we can 
describe correctly both the probability distribution of 
measurement readouts and the final state of the measured system. 

Thus, the influence of the environment on the system of interest 
may be given in terms of information. Therefore, the RPI approach 
is in fact an {\em information approach} to the theory of 
continuous quantum measurements. This indicates the
fundamental character of the approach. 

The information approach in the quantum theory of measurement is 
not novel. The very first quantum theory of (instantaneous) 
measurements based on the 
von Neumann reduction postulate may be considered as an example 
of an information approach. In this theory both the probability 
distribution for measurement outputs and the final state of the 
system may be found if we know what information the measurement 
supplies. For example, for the measurement of an observable with 
a discrete spectrum we need to know what eigenvalue is obtained 
as the measurement output. Given this information, we can 
determine, with the help of the corresponding projector, both the 
probability of the measurement output and the final state of the 
system. 

In the RPI theory of continuous quantum measurements the 
information principle obtains one more realization having a 
rich structure and wide range of applications.  It has been 
shown above that the RPI theory is, in contrast to the von 
Neumann postulate, a natural part of (the path-integral 
version of) quantum mechanics. 

\section{Conclusion}

The restricted-path-integral (RPI) approach to continuous quantum 
measurements enables one to describe the influence of the 
measurement on the measured system without explicitly 
considering any model of the measuring device (environment). 
Instead of the model, one needs to know the information 
supplied by the measurement. 

It is remarkable that knowledge of this information is sufficient for 
correctly accounting for the influence of the measuring environment 
on the system. This feature enables one to derive the RPI theory 
of continuous measurements from the path-integral version of 
quantum mechanics without any additional postulates. Besides, 
this shows that the RPI approach is an 
information approach to continuous measurements just as the 
von Neumann projection theory is an information approach to 
instantaneous measurements. 

The information character of  the RPI theory indicates clearly  that 
it reveals deep internal qualities of quantum mechanics. This theory 
may be used to investigate further the relations between quantum 
and classical physics, between the quantum world and its classical 
appearance. 

\vspace{0.5cm}
\centerline{\bf ACKNOWLEDGMENTS}

This work was supported in part by the Deutsche 
Forschungsgemeinschaft.


\end{document}